\pdfoutput=1
\documentclass[conference]{IEEEtran}
\IEEEoverridecommandlockouts
\usepackage[utf8]{inputenc}
\usepackage[T1]{fontenc}
\usepackage{newtxtext,newtxmath}

\usepackage[english]{babel}
\usepackage{amsmath,amssymb}
\usepackage{graphicx}
\usepackage{booktabs}
\usepackage{cite}
\usepackage{xcolor}
\usepackage[hidelinks]{hyperref}

\begin{document}
\title{From Tokens to Pixels: Training-Free Guided Upsampling of Frozen CARL Features for Hyperspectral Imaging}
\author{\IEEEauthorblockN{Dominik Fromme, Ellen Markowski, Tim Streckert and Jörg Thiem}
\IEEEauthorblockA{University of Applied Sciences and Arts Dortmund\\
Dortmund, Germany\\
\{dominik.fromme, ellen.markowski, tim.streckert, joerg.thiem\}@fh-dortmund.de\\
}}
\pagestyle{plain}
\maketitle
\thispagestyle{plain}

\begin{abstract}
Camera-agnostic foundation models such as CARL map spectral images from arbitrary sensors into a shared feature space, but deliver their features only on a token grid at one eighth of the input resolution. The usual bilinear interpolation blurs material boundaries over about one token and rounds off corners. We investigate how much spatial accuracy image-guided upsampling recovers for frozen CARL features without training the encoder or the upsampler. We compare joint bilateral upsampling (JBU), a guided filter and the pretrained AnyUp, each with the hyperspectral image or an RGB rendering as the guide. A hyperspectral ColorChecker dataset acquired specifically for this study, with model-independent reference geometry measured to sub-pixel precision, serves as the evaluation benchmark. JBU with the hyperspectral guide raises the Boundary IoU from 13 to 67~\% and reduces the effective edge width of the features from 38 to 4~px, close to the resolution of the raw spectra. The effect occurs in all 24 patches of the ColorChecker and replicates on an additional recording. On placenta and oral cavity images, the feature edges also become markedly qualitatively sharper. For sharp, high-contrast boundaries, JBU is the most accurate and most economical method. On tissue with softer transitions, the pretrained AnyUp yields qualitatively sharper contours. An analysis of 78 simulated band configurations shows that the gain persists in every configuration as long as the head matches the configuration. Image-guided upsampling is thus a simple, training-free step for applications that use CARL frozen with linear heads.
\end{abstract}

\begin{IEEEkeywords}
Hyperspectral imaging, foundation models, feature upsampling, joint bilateral upsampling, Boundary IoU
\end{IEEEkeywords}

\begin{figure*}[!t]
\centering
\begin{tabular}{@{}l@{}}
\textbf{(a)}\\ \includegraphics{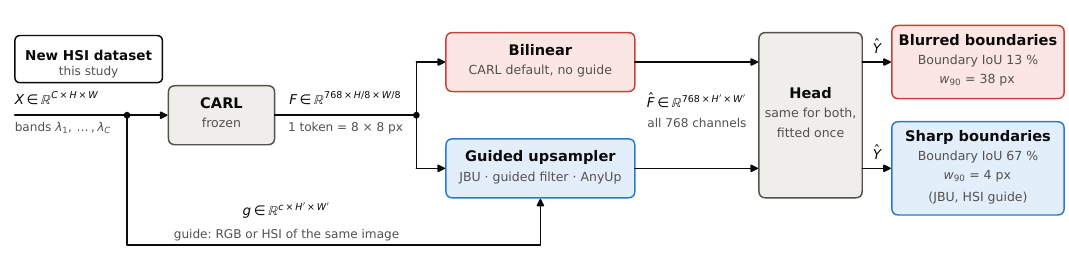}\\[1mm]
\textbf{(b)}\\ \includegraphics{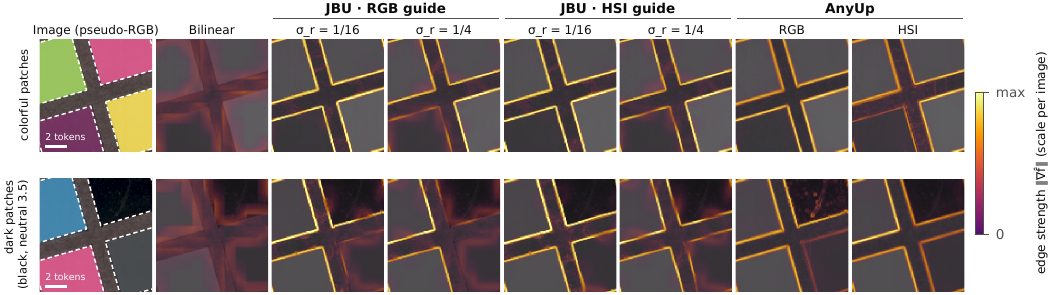}
\end{tabular}
\caption{(a) Pipeline. CARL stays frozen and returns one feature per token of 8~×~8 input pixels. Only the upsampling step is compared. Bilinear interpolation, the default of CARL, uses no guide. JBU, the guided filter and AnyUp take the guide $g$ from the same image. All upsamplers bring the 768 feature channels to the output grid $H' \times W'$ and feed the same head, which is fitted once on tokens. Numbers as in Table~\ref{tab:main} (prototype head). (b) Where do the CARL features change? Sobel gradient magnitude of the unit-normalized features at two patch junctions of the recording rotated by 16°, overlaid on the dimmed image (color scale per row, gamma 0.6, as in Fig.~\ref{fig:medical}). White dashed: measured patch outline. One token covers 50~×~50 native pixels. Top: colorful patches. Bottom: dark patches with low visible contrast to the frame. Bilinear interpolation blurs the edges over about one token. JBU places the edges on the outline but keeps a fringe of about one token at low-contrast boundaries, which widens with $\sigma_r = s_g/4$. With the RGB guide, the fringe partly replaces the edge at Neutral 3.5. AnyUp also hits the outline, with even narrower edges. With the RGB guide, it additionally traces fine specks on the black patch. $\sigma_r = s_g/16$ is the value selected on the respective other recording.}
\label{fig:teaser}
\end{figure*}

\section{Introduction}
Foundation models for spectral images aim to make multispectral and hyperspectral data usable without task-specific training~\cite{carl,spectralgpt,hypersigma,dofa}. CARL~\cite{carl} is explicitly designed for changing cameras. It combines a wavelength-aware spectral encoder, which aggregates arbitrary band configurations, with a vision transformer for spatial context. Recordings from different sensors thus end up in a shared feature space and can be analyzed with simple heads such as linear ones, but at the price of spatial resolution.

Like a vision transformer, CARL splits the input image into non-overlapping
patches of $8 \times 8$ pixels and represents each patch by a single token.
Its features therefore lie on a token grid at one eighth of the input
resolution. How many camera pixels a token covers depends on how the image is resampled before encoding. In our main setting, one token covers
$50 \times 50$ native camera pixels (Fig.~\ref{fig:grids}). For dense predictions, the reference CARL implementation bilinearly interpolates this grid. Because each token mixes the properties of all materials within its footprint and bilinear weights depend only on distance, material boundaries are blurred over about one token and right-angled corners become rounded (Fig.~\ref{fig:teaser}b). Learnable upsamplers for vision encoders~\cite{featup,loftup,jafar} address this problem but usually have to be trained for each encoder.

In medical hyperspectral imaging, accurate tissue boundaries are particularly important. Examples include tumor margins, blood vessel contours and transitions between perfused and ischemic tissue. Skin and tissue models that assign parameters such as perfusion, optical or thermal properties to individual regions likewise require precise boundaries. At the same time, medical datasets are small and laborious to annotate. A frozen encoder with a linear head is attractive in this setting because it requires few labels, reduces the risk of overfitting and remains interpretable. The key question is how much edge accuracy can be recovered without additional training.

Image-guided upsamplers from image processing~\cite{jbu,gf} weight neighboring tokens by how similar they are to a high-resolution guide image at the target pixel. For hyperspectral data, the input image itself can always serve as such a guide. It is much finer than the token grid and carries spectral information beyond RGB. Fig.~\ref{fig:teaser}a summarizes the pipeline and the comparison. This work contributes:
\begin{itemize}
\item a systematic evaluation of training-free, image-guided upsamplers (JBU, guided filter, AnyUp) for frozen CARL features, each with a hyperspectral and an RGB guide and under a pre-specified protocol
\item a hyperspectral dataset of a ColorChecker with model-independent, sub-pixel reference geometry, together with the contamination width $w_{90}$ as a measure of the effective resolution of the features
\item a qualitative transfer to placenta and oral cavity images
\item a CARL-specific analysis of robustness to wavelength errors and other band sets
\end{itemize}
Code, result tables and access to the ColorChecker data are provided at \url{https://gitlab.fb10.fh-dortmund.de/fb10-opensource/CARL-tokens-to-pixels}.

\section{Related Work}
\subsection{Foundation Models for Spectral Images}
Besides CARL~\cite{carl}, SpectralGPT~\cite{spectralgpt} for multispectral Sentinel-2 data, HyperSIGMA~\cite{hypersigma} for hyperspectral data and DOFA~\cite{dofa} for multiple Earth observation sensors have been proposed. Only DOFA and CARL explicitly condition on the wavelengths and thus accept changing band configurations. All operate on patch tokens and deliver features at reduced spatial resolution. In medical hyperspectral imaging, public data are scarce~\cite{patzke}. Segmentation networks have so far mostly been trained task-specifically~\cite{seidlitz,heipor}.

\subsection{Feature Upsampling}
Classical image-guided methods are joint bilateral upsampling~\cite{jbu} and the guided filter~\cite{gf}. For the features of vision encoders, learnable upsamplers such as FeatUp~\cite{featup}, LoftUp~\cite{loftup} and JAFAR~\cite{jafar} have been proposed. AnyUp~\cite{anyup} and NAF~\cite{naf} are encoder-agnostic and can be applied to new features without retraining. These works evaluate almost exclusively RGB encoders and RGB guides.

\subsection{Edge-Preserving Filters in Hyperspectral Imaging}
In spectral–spatial classification, edge-preserving filtering~\cite{epf} smooths the probability maps of a pixel-wise classifier with a guided or bilateral filter. The first principal components of the hyperspectral image serve as the guide. In contrast, we filter not the output of a classifier but the features of a frozen encoder before the head. We are not aware of any work that systematically examines how much edge fidelity training-free, image-guided upsampling recovers for a camera-agnostic hyperspectral encoder and whether the hyperspectral image is a better guide than RGB.

\section{Method}
\subsection{Problem Statement}
CARL maps a spectral image $X \in \mathbb{R}^{C \times H \times W}$ with band wavelengths $\lambda_1,\dots,\lambda_C$ to a token map $F \in \mathbb{R}^{D \times H/8 \times W/8}$ with $D = 768$ channels. Each token feature $f_t \in \mathbb{R}^D$ belongs to a footprint $\Omega_t$ of $8 \times 8$ input pixels with center $c_t$. A dense prediction requires a feature $\hat f(x) \in \mathbb{R}^D$ at every pixel $x$ of an output grid of $H' \times W'$ pixels, which is finer than the token grid but otherwise independent of the input size $H \times W$. Here, the output grid is the evaluation grid of $5 \times 5$-pixel blocks, so one token covers $10 \times 10$ output pixels (Fig.~\ref{fig:grids}a–d). Positions $x$ and $c_t$ are given in token units. Except for the guided filter, all methods considered compute a convex combination of the tokens,
\begin{equation}
\hat f(x) = \sum_t \alpha(x,t)\, f_t, \qquad \alpha(x,t) \ge 0, \qquad \sum_t \alpha(x,t) = 1,
\label{eq:convex}
\end{equation}
and differ only in the weights. Nearest selects the token whose footprint contains $x$. Bilinear interpolation uses tent weights based on the distance to the token centers. Neither set of weights depends on the image content, so the feature edges lie on the token scale. Because the weights form a convex combination, $\hat f(x)$ stays within the range of the tokens and their semantics are preserved. For a linear head, it also makes no difference whether features or logits are upsampled. Every method upsamples all $D$ feature channels. The guide image only determines the weights.

\subsection{Image-Guided Upsamplers}
Joint bilateral upsampling~\cite{jbu} computes each output pixel from a local neighborhood of tokens. We use a $5 \times 5$ window centered on the token containing the output pixel, extending two tokens in each direction. Within this window, token contributions are weighted by spatial distance and guide similarity,
\begin{equation}
\alpha(x,t) \propto \exp\!\left(-\frac{\lVert x - c_t \rVert^2}{2\sigma_s^2} - \frac{\lVert g(x) - \bar g_t \rVert^2}{2\sigma_r^2}\right),
\label{eq:jbu}
\end{equation}
where $c_t$ is the center of token $t$, $g(x)$ is the guide at output pixel $x$ and $\bar g_t$ is the mean guide value over the footprint $\Omega_t$ of token $t$. As in~\cite{jbu}, the spatial bandwidth is set to $\sigma_s = 0.5$~token. The guides are scaled to an RMS spread $s_g = 1$, so the bandwidth $\sigma_r$ has the same scale for the RGB and the hyperspectral guide.

\begin{figure*}[!t]
\centering
\includegraphics{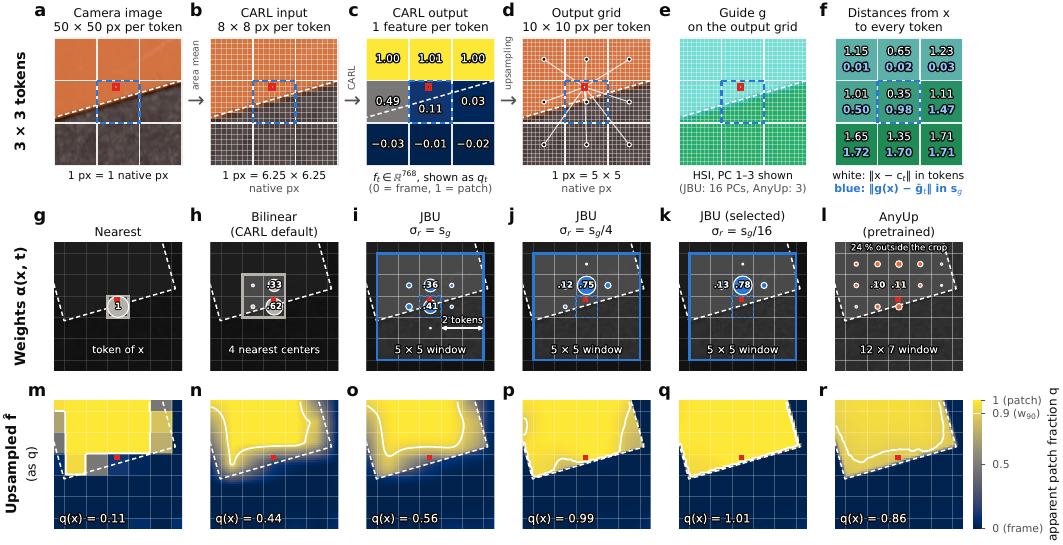}
\caption{Grids and upsampling weights for one output pixel $x$ (red) just inside the lower edge of the orange patch in the rotated recording. Native px are the pixels of the camera. (a–d) The same $3 \times 3$ tokens at each step of the pipeline. The token that contains $x$ is outlined in blue. One token covers $50 \times 50$ native px (a). For CARL, the image is resampled by area averaging to $8 \times 8$ input px per token, so one input px corresponds to $6.25 \times 6.25$ native px (b). CARL returns one feature $f_t \in \mathbb{R}^{768}$ per token (c). The upsamplers compute $\hat f(x)$ on the output grid with $10 \times 10$ px per token, so one output px corresponds to $5 \times 5$ native px (d). The guide and the evaluation use this grid as well. Features are shown as apparent patch fraction $q$ (0 = frame, 1 = patch, Section~\ref{sec:metrics}). Although $x$ lies on the patch, its token mainly covers the frame ($q_t = 0.11$). (e) Hyperspectral guide on the output grid, shown by its first three principal components. JBU uses the first 16, AnyUp the first three. (f) The two distances that enter JBU (Eq.~\ref{eq:jbu}) for every token. White: distance between $x$ and the token center $c_t$ (lines in d) in token units. Blue: distance between the guide at $x$ and the mean guide of the token in units of $s_g$. The token of $x$ is closer in position (0.35 compared with 0.65 for the token above). The token above, however, matches the guide (0.02 compared with 0.98). (g–l) Weights $\alpha(x,t)$ as disks with area proportional to the weight. Only tokens in the framed region can contribute. These are the token of $x$ (nearest), the four nearest token centers (bilinear) and the $5 \times 5$ window centered on the token of $x$ (JBU). AnyUp attends to a window that spans $\pm 10$~\% of the image height and width around $x$, here $12 \times 7$ tokens. It extends beyond the crop and 24~\% of the weight lies outside. (m–r) Upsampled features as $q$ with the contour $q = 0.9$ (solid) and the measured outline (white dashed). Their distance, averaged along a side, corresponds to the contamination width $w_{90}$. Bilinear weights depend only on the position. The smaller $\sigma_r$, the more weight JBU gives to the token above. The edge of $\hat f$ then moves to the measured outline. AnyUp spreads its weight over many tokens, mostly on the patch, and reaches $q(x) = 0.86$.}
\label{fig:grids}
\end{figure*}

As in edge-preserving filtering~\cite{epf}, the guided filter~\cite{gf} is applied after bilinear upsampling. Let $p(x) \in \mathbb{R}^D$ be the bilinearly upsampled feature and $g(x) \in \mathbb{R}^3$ the guide. In every window $\omega_k$ of radius one token centered at pixel $k$, the filter fits a linear model $p \approx a_k^\top g + b_k$ with $a_k \in \mathbb{R}^{3 \times D}$ and $b_k \in \mathbb{R}^D$,
\begin{equation}
\begin{aligned}
a_k &= (\Sigma_k + \varepsilon I)^{-1}\,\mathrm{cov}_k(g, p), \qquad b_k = \bar p_k - a_k^\top \bar g_k,\\
\hat f(x) &= \bar a_x^\top g(x) + \bar b_x,
\end{aligned}
\label{eq:gf}
\end{equation}
where $\Sigma_k \in \mathbb{R}^{3 \times 3}$ is the covariance of the guide in $\omega_k$, $\mathrm{cov}_k(g, p) \in \mathbb{R}^{3 \times D}$ the covariance between guide and feature, $\bar g_k$ and $\bar p_k$ their means. $\bar a_x$ and $\bar b_x$ are the averages of $a_k$ and $b_k$ over all windows that contain $x$. The regularization $\varepsilon$ limits how strongly edges of the guide pass into the features. The guided filter is the only method that is not a convex combination. AnyUp~\cite{anyup} computes the weights with a pretrained, encoder-agnostic window attention between guide pixels and tokens. We use the published model without retraining. Fig.~\ref{fig:grids} shows the grids involved and the weights of bilinear interpolation, JBU and AnyUp for one pixel.

The guide is either an RGB rendering of the cube (440–500, 510–570 and 600–680~nm) or the hyperspectral image itself. JBU uses its first 16 principal components. AnyUp receives the first three principal components as a false-color image, because it is pretrained on three-channel RGB images and more channels would require retraining. As in edge-preserving filtering~\cite{epf}, the guided filter also uses the first three principal components and is evaluated only with this guide. All guides are formed on the evaluation grid of 5~×~5-pixel blocks, ten times finer than the 50-px token grid.

\section{Experimental Setup}
\subsection{Dataset}
\begin{figure*}[!t]
\centering
\includegraphics{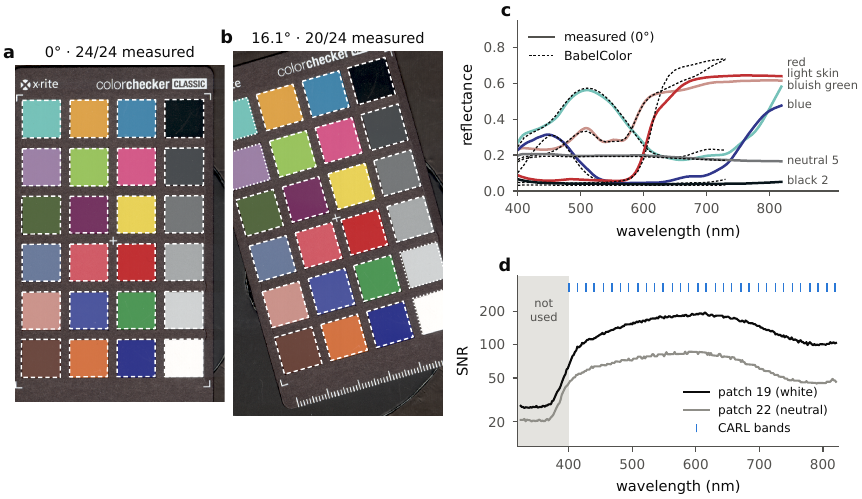}
\caption{Dataset and reference. (a, b) Both recordings as pseudo-RGB with measured patch outlines (white dashed) and model-placed, cut-off patches (white dotted). (c) Measured reflectance spectra of selected patches (mean over the patch interior, upright recording) and, dashed, the BabelColor average of the same patches~\protect\cite{babel}, tabulated from 380 to 730~nm. (d) SNR per band in the white and the neutral 5 patch. Bands below 400~nm (gray) are not used. Blue ticks mark the 32 CARL bands.}
\label{fig:data}
\end{figure*}

We acquired the recordings with the setup described in~\cite{patzke}. It consists of a Resonon Pika UV-505 push-broom camera (CMOS, 12~bit, 1500 pixels across track) with a Jenoptik UV-VIS-IR 60~mm 1:4 APO Macro lens (apochromatic for 315–1100~nm) on a Resonon benchtop system whose linear stage moves the sample under the camera, inside an enclosure lined with absorbing foil. Illumination is provided by a halogen line light fed via an optical fiber. The UVA fluorescent tube of the setup (UV lamp in the following) was additionally switched on only for the third recording. With 4-fold spectral binning, the camera delivers 270 bands between 325 and 821~nm (spacing 1.8–1.9~nm, full width at half maximum 2.8~nm). Patzke et al.~\cite{patzke} use the 255 bands within 330–800~nm. Exposure time and line rate are 222~ms and 4.5~Hz (232~ms and 4.3~Hz for the third recording) at gain 0. The working distance is about 45~cm. The acquisition software converts raw values $I$ into reflectance using white and dark references $W$ and $D$,
\begin{equation}
R(\lambda) = \frac{I(\lambda) - D(\lambda)}{W(\lambda) - D(\lambda)}.
\end{equation}
An X-Rite ColorChecker Classic Mini with 24 patches was recorded in two orientations, upright (2429~×~1500~px) and rotated by 16.1° (2600~×~1500~px). In the rotated recording, the image border cuts four patches (Fig.~\ref{fig:data}). A token of the main setting covers 50~×~50~px, about one seventh of the patch pitch of 340~px. The frame gap between two patches is only 1.4 tokens wide at about 70~px. A third, upright recording with the UV lamp additionally switched on (2700~×~1500~px) was evaluated only after the main analysis, as a replication and as a test of the UV range (Section~\ref{sec:guide}).

\subsection{Reference Geometry}
The reference does not depend on any model. Each fully visible patch is measured at native resolution. From pseudo-RGB and NIR, we form a normalized contrast that is 0 on the frame and 1 inside the patch. Its 0.5 isoline defines the edge. We fit a total-least-squares line to each side and take the intersections of adjacent lines as the corners. The patch identities follow from comparing the measured spectra with the BabelColor reference spectra~\cite{babel} over all rotations and reflections of the chart. All 24 patches are measured in the upright recording and 20 in the rotated one. Each side is located to within 0.8~px RMS, far below the 10-px band of the Boundary IoU. An ideal chart model, i.e., a homography that maps a perfect patch grid into the image, matches the measured corners only to 2.2–2.4~px RMS. The largest deviations occur in both recordings at the same four low-contrast patches (Dark Skin, Black, Foliage, Neutral 3.5), which are measured about 3.5–7~px larger than the median of the other patches (mean side lengths 265–275~px). For three of them, the reference edge follows the NIR channel and lies about 3~px further out than in the visible range. We therefore use this chart model, i.e., the fitted homography of the ideal grid, only where nothing can be measured. It defines the outer boundary of the evaluated frame region, half a frame gap beyond the outer patches. Everything outside is ignored. It also places the four patches in the rotated recording that are cut by the image border. The classes are the 24 patches and the frame. The radiometry is consistent. The median color difference $\Delta E_{00}$ is 0.2 between the recordings and 1.7–1.8 to the BabelColor average (30 charts, measured before the 2014 reformulation of the colors).

\subsection{Features and Heads}
We use the public CARL SSL checkpoint~\cite{carl} frozen. CARL receives 32 linearly spaced bands between 400 and 820~nm together with their wavelengths, normalized per image to zero mean and unit standard deviation. Bands below 400~nm are not used because their SNR is clearly below that of the visible range (Fig.~\ref{fig:data}d) and CARL was pretrained on sensors starting at about 420~nm. The input is area-averaged exactly so that a token covers 50~×~50 native pixels. We additionally examine 20 and 25~px. We evaluate the spatial token map (768 channels), which CARL's segmentation head also reads, on a grid of 5~×~5-pixel blocks.

Tokens whose footprint lies at least 95~\% within one class form the support of that class. Two heads are fitted once on these tokens and never see an upsampled feature, so differences are due to the upsampler alone. The first is a training-free prototype head (class means of the support tokens, assignment by the smallest angle). The second is a linear probe (multinomial logistic regression, corresponding to CARL's $1 \times 1$ convolutional head). The prototype rule on the raw spectra with all 229 bands between 400 and 820~nm serves as an upper reference. In this work, training-free refers to the upsampling. CARL stays frozen, JBU and the guided filter have no learned parameters and AnyUp is used as published. Only the heads are fitted once on labeled support tokens.

\subsection{Metrics}
\label{sec:metrics}
The Boundary IoU~\cite{biou} assesses class membership only within a band of $d = 10$ native pixels around the contours of reference and prediction, the Corner IoU within circles of 25~px radius around the four reference corners of a patch. We measure the effective resolution of the features directly on the features. For a patch with prototype $\pi_k$ next to the frame with prototype $\pi_f$,
\begin{equation}
q(s) = \frac{\langle \hat f(s) - \pi_f,\; \pi_k - \pi_f \rangle}{\lVert \pi_k - \pi_f \rVert^2}
\label{eq:q}
\end{equation}
is the apparent patch fraction at distance $s$ from the reference edge ($s > 0$ inside the patch), averaged along the side. Since Eq.~\ref{eq:q} is normalized to the prototypes, which are means over all support tokens, single tokens can yield $q$ slightly below 0 or above 1. Edge tokens with up to 5~\% frame in their footprint add to this, as they shift $\pi_k$ slightly toward the frame. We average only over the central 60~\% of each side. Each profile thus stays about 55~px, i.e., more than one token, away from the adjacent sides and captures only the one edge. The corners are assessed by the Corner IoU. The contamination width $w_{90}$ is the distance into the patch beyond which $q \ge 0.9$ holds, i.e., less than 10~\% frame is mixed in, taken as the median over the four sides.

\subsection{Protocol and Statistics}
All settings were fixed before the first upsampling result. The only later change, also made before any result, replaced 100-px tokens in the token sweep with 20~px, because 100-px tokens are wider than the frame gap. The two parameters without a transferable default, $\sigma_r \in \{1/16, \dots, 1\} \cdot s_g$ and $\varepsilon \in \{0.01^2, \dots, 0.2^2\}$, are selected across recordings. The best value of one recording is used on the other. Thus, no reported number is optimized on its own labels.

The statistical unit is the physical patch ($n \le 24$). For each patch, the difference to bilinear interpolation is averaged over the recordings. We report the median of these differences with a 95~\% cluster bootstrap confidence interval (10\,000 resamples~\cite{efron}) and test them with the Wilcoxon signed-rank test. The primary comparison, fixed in advance, is JBU with the hyperspectral guide versus bilinear in Boundary IoU (spatial features, prototype head, $\alpha = 0.05$). All other comparisons are Holm-corrected within their analysis~\cite{holm}.

\subsection{Medical Images}
Our qualitative evaluation uses two public datasets. The first contains placental images acquired with a Senop HSC-2 camera mounted on an operating microscope (1024~×~1024~pixels)~\cite{placenta}. The second, ODSI-DB~\cite{odsi}, contains images of teeth, oral mucosa and lips acquired using either a Specim IQ (204 bands, 400–1000~nm) or a Nuance EX (51 bands, 450–950~nm). For the placenta, we use the 35 bands in 515–698~nm that all images contain, excluding the bands around 640~nm with a known sensor anomaly. For ODSI-DB, we use the common range of 450–950~nm and every third Specim band. CARL receives both at half resolution, so a token covers 16~×~16 native pixels. We examine the edge maps of the upsampled features themselves, without a head and without training. They show the Sobel gradient magnitude of the unit-normalized features, as in Fig.~\ref{fig:teaser}b. We deliberately keep this part qualitative, because the accuracy of the annotated boundaries in both datasets is unknown and boundary metrics would partly measure label uncertainty.

\subsection{Implementation}
The experiments were implemented in Python with PyTorch and run on an NVIDIA H100 GPU. The analysis and figure code was developed with the AI coding assistant Claude Code (Claude Opus 5.5, Anthropic) under the authors' direction and checked by them. All numbers and figures are computed by this code from the recorded data. Code and notebooks are published for verification.

\section{Results}
\subsection{Edges and Corners}
\begin{figure*}[!t]
\centering
\includegraphics{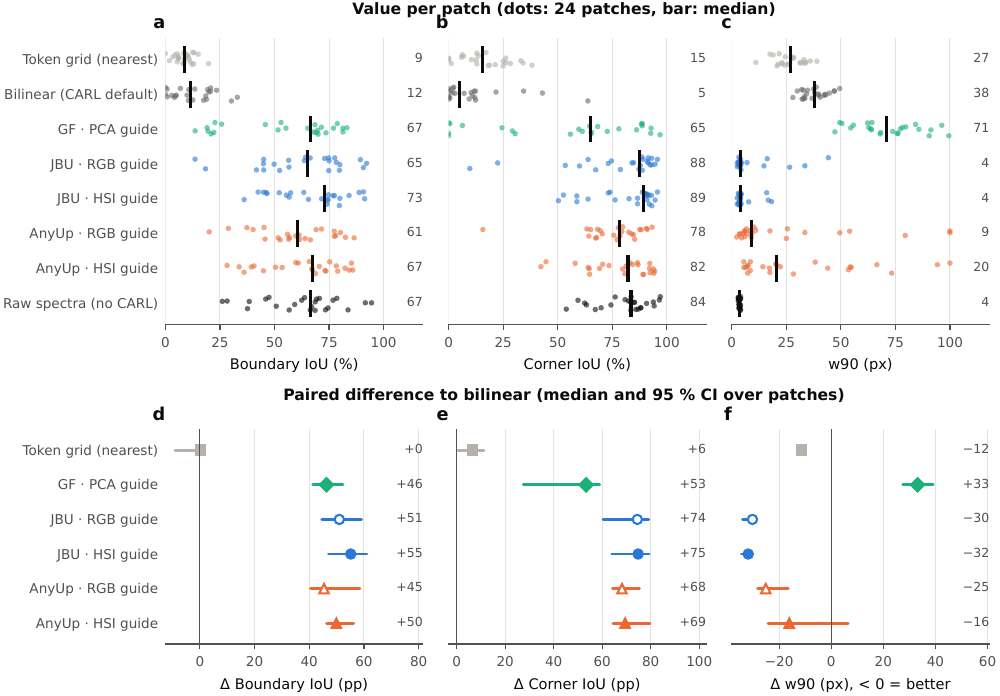}
\caption{Main result (spatial CARL features, prototype head, 50-px tokens). (a–c) Value of each measured patch (dots, mean of both recordings) and median (bar) for Boundary IoU, Corner IoU and $w_{90}$, with $w_{90}$ censored at 100~px. (d–f) Median of the paired difference to bilinear interpolation with 95~\% cluster bootstrap CI. For $w_{90}$, a negative value means narrower edges.}
\label{fig:main}
\end{figure*}

\begin{figure*}[!t]
\centering
\includegraphics{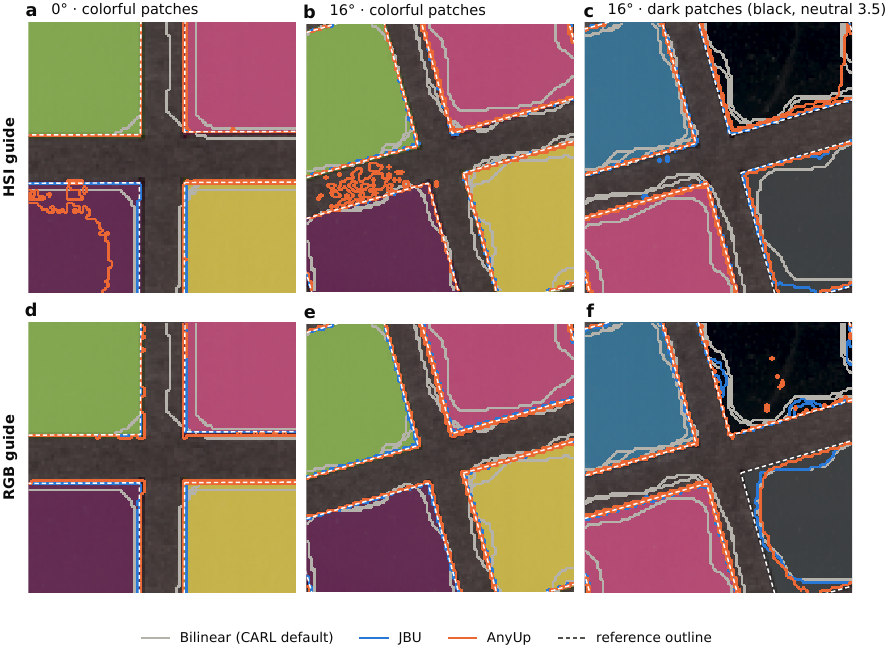}
\caption{Decision boundaries of the prototype head at three patch junctions, top with the hyperspectral guide (a–c), bottom with the RGB guide (d–f). White dashed: measured outline. Bilinear rounds off corners and pulls edges into the patch. JBU follows the outline with both guides. At the dark patches, JBU with the RGB guide deviates from the outline (f), with the hyperspectral guide it does not (c).}
\label{fig:boundaries}
\end{figure*}

JBU with the hyperspectral guide raises the Boundary IoU over bilinear interpolation by a median of 55~pp (95~\% CI 47–61~pp, $p \approx 10^{-7}$), in all 24 patches and equally in both recordings. On the third recording, evaluated later, the gain is +52~pp, also in all 24 patches. All other guided methods also clearly improve Boundary IoU and Corner IoU, for both heads and both feature maps (Holm $p < 0.001$, Table~\ref{tab:main}, Fig.~\ref{fig:main}). The difference is most pronounced at the corners. Bilinear interpolation rounds them off so strongly that it performs worse than the raw token grid (Fig.~\ref{fig:boundaries}). With JBU, the CARL features reach the level of the raw spectra in Boundary IoU.

\begin{table*}[!t]
\centering
\footnotesize
\caption{Results on the ColorChecker (spatial CARL features, mean of both recordings, 50-px tokens). IoU values in \%, mean over measured patches. Median $w_{90}$ in native px.}
\label{tab:main}
\begin{tabular}{lccccc}
\toprule
Method & Boundary IoU (prototype) & Boundary IoU (linear) & Corner IoU (prototype) & Patch IoU (prototype) & $w_{90}$ (px) \\
\midrule
Token grid (nearest) & 8.9 & 9.9 & 16.5 & 70.7 & 27.9 \\
Bilinear (CARL default) & 13.3 & 18.4 & 10.0 & 78.1 & 38.2 \\
Guided filter, PCA guide & 56.6 & 66.6 & 55.9 & 92.9 & 71.0 \\
JBU, RGB guide & 62.8 & 67.1 & 77.0 & 94.9 & 3.8 \\
\textbf{JBU, HSI guide} & \textbf{67.2} & \textbf{70.3} & \textbf{81.4} & \textbf{96.2} & \textbf{3.8} \\
AnyUp, RGB guide & 60.4 & 64.3 & 75.7 & 95.3 & 9.2 \\
AnyUp, HSI guide & 63.6 & 69.2 & 78.3 & 94.9 & 22.6 \\
Raw spectra (no CARL) & 63.2 & -- & 80.4 & 95.7 & 3.2 \\
\bottomrule
\end{tabular}
\end{table*}

\subsection{Effective Resolution of the Features}
\begin{figure}[!t]
\centering
\includegraphics[width=\columnwidth]{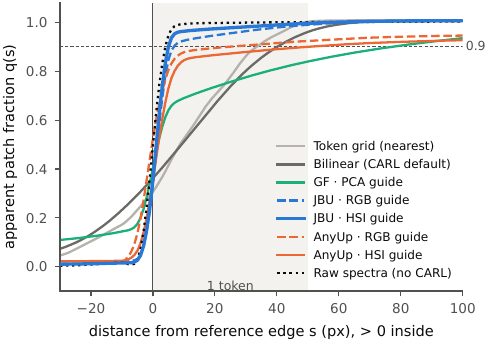}
\caption{Apparent patch fraction $q(s)$ over all measured patch edges of both recordings. Gray background: one token (50~px). Dashed: threshold 0.9 for $w_{90}$.}
\label{fig:profiles}
\end{figure}

After bilinear upsampling, frame information reaches a median of 38~px, almost one token, into a patch before less than 10~\% is mixed in. With JBU, it is 4~px, barely more than for the raw spectra (Fig.~\ref{fig:profiles}). The profiles show two further failure patterns. The guided filter places the edge correctly but leaves a long tail inside the patch, because it corrects the bilinear ramp, which is about two tokens wide, only in windows that contain the edge. AnyUp yields sharp edges but, for some dark patches, still mixes in frame far inside the patch. Edge maps show only where the features change, not how pure they are. In Fig.~\ref{fig:teaser}b, the edges of AnyUp even appear the narrowest, whereas JBU keeps a faint fringe at low-contrast boundaries. Only the profiles capture the residual mixing inside the patch. The findings do not depend on the threshold of 0.9 (Appendix~\ref{app:edge}).

\subsection{Token Size and Parameter Selection}
\begin{figure*}[!t]
\centering
\includegraphics{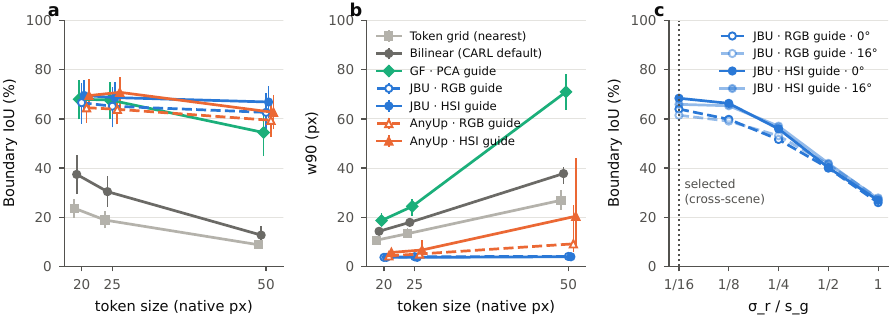}
\caption{(a) Boundary IoU and (b) $w_{90}$ as a function of the token size. Dots: mean ($w_{90}$: median) over patches, error bars: 95~\% bootstrap CI. (c) Boundary IoU of JBU as a function of the bandwidth $\sigma_r$ (in units of the RMS spread $s_g$ of the guide) for both recordings. Dashed: the value selected across recordings.}
\label{fig:footprint}
\end{figure*}

The gain grows with the token size, because bilinear interpolation deteriorates with larger tokens while JBU remains nearly constant ($w_{90} \approx 4$~px, Fig.~\ref{fig:footprint}a, b). With an image guide, the effective resolution of the features is thus decoupled from the token size in this range. The Boundary IoU of JBU decreases monotonically with $\sigma_r$. The curves of both recordings nearly coincide (Fig.~\ref{fig:footprint}c), so the selection across recordings finds the best value. For the guided filter, it costs about 4~pp in one recording. The spatial bandwidth $\sigma_s$ is uncritical (Appendix~\ref{app:sens}).

\subsection{Hyperspectral versus RGB Guide}
\label{sec:guide}
Across all patches, the hyperspectral guide is not better than the RGB guide, neither for JBU nor for AnyUp (Holm $p \ge 0.18$). For most patches, the RGB image alone separates patch and frame. For JBU, the gain is concentrated on a few patches with low visible contrast to the dark frame (Neutral 3.5, Black, Dark Skin, Foliage), with up to +29~pp Boundary IoU. Its size follows the spectral angle between patch and frame, not the color difference $\Delta E_{00}$ (Appendix~\ref{app:h2}). With reference edges from the visible range alone, this pattern and all main results remain, although the gain on Foliage becomes smaller. The hyperspectral image thus helps exactly where two materials differ spectrally but look similar in the RGB image. For AnyUp, the differences between the guides scatter in both directions and follow the spectral angle only in $w_{90}$.

With the third recording, we test whether the bands below 400~nm, which an RGB image does not capture, provide the guide with additional edges. The camera records these bands in every recording. The UV lamp was meant to strengthen their signal. This does not happen. The reflectance equals that of the recording without the lamp (median $\Delta E_{00}$ 0.3). The SNR below 400~nm also remains unchanged. Accordingly, a guide from the principal components of 325–820~nm is not better than one from 400–820~nm (JBU and AnyUp $-0.1$~pp, Holm $p \ge 0.3$, Table~\ref{tab:uv}). The principal component analysis suppresses the UV only partly. The UV bands carry 3~\% of the variance and barely enter the three components for AnyUp, but the 16 components for JBU retain 99.9~\% of their variance. Weighting the UV more strongly in a post hoc test makes the guide worse. With a UV-only guide, JBU loses a median of 17~pp Boundary IoU. With a guide that is half UV, it still loses 4~pp (Holm $p = 0.02$). The UV bands are blurred along the slit and therefore carry less edge information than the visible range. Whether UV information can improve the guide thus remains open.

\subsection{Robustness and Cost}
With heads from the respective other recording, the Boundary IoU drops by at most 4~pp. The ranking of the methods is preserved. The spectral feature map and the linear probe give the same picture (Tables~\ref{tab:spectral} and~\ref{tab:cross}). For a feature map with 768 channels at 520~×~300~px, JBU takes about 40~ms. AnyUp needs about twice the time and memory, the guided filter almost three times the time (Table~\ref{tab:runtime}). The guide does not change this. JBU takes equally long with three RGB channels and with 16 principal components, because the cost is dominated by the 768 feature channels. AnyUp receives a three-channel image in both cases. The principal component analysis for the guide is not included.

\subsection{Medical Images}
\begin{figure*}[!t]
\centering
\includegraphics{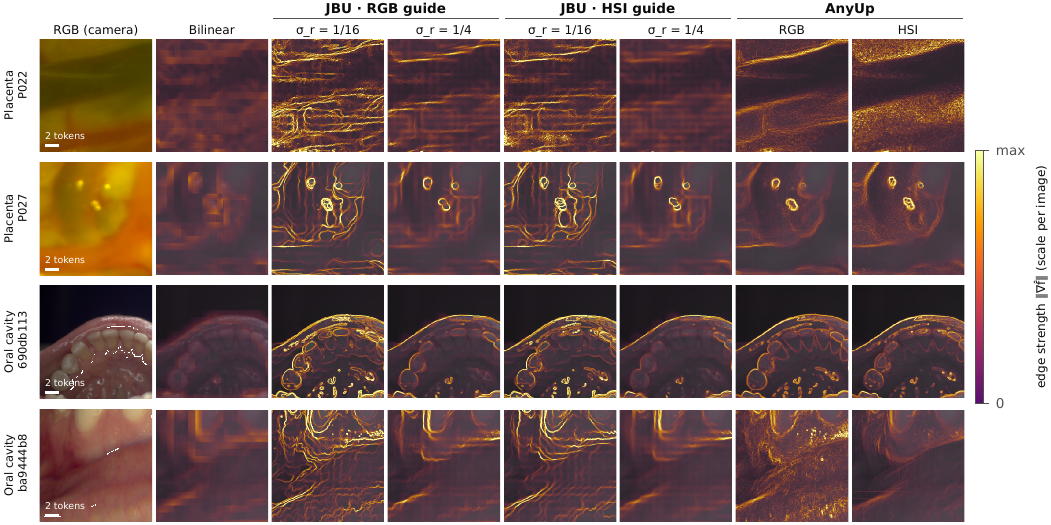}
\caption{Edge maps of the upsampled CARL features on placenta and oral cavity images, without a head and without training, overlaid on the dimmed image (color scale per image, gamma 0.6 emphasizes weak edges). Crops of 256~×~256~px (16~×~16 tokens). Placenta as the unprocessed RGB image, oral cavity with the annotated class boundaries (white) for orientation. JBU with the RGB or hyperspectral guide (16 principal components) and $\sigma_r = s_g/16$ or $s_g/4$. AnyUp with the RGB image or the hyperspectral false-color image (principal components 1–3) as the guide. Images adapted from the placenta dataset~\protect\cite{placenta} and ODSI-DB~\protect\cite{odsi}. This figure is licensed under CC BY-NC-SA 4.0 and is not covered by the license of this article.}
\label{fig:medical}
\end{figure*}

On tissue, the same pattern appears qualitatively as on the ColorChecker (Fig.~\ref{fig:medical}, cf. Fig.~\ref{fig:teaser}b). Upsampled bilinearly, the features form a blurred mosaic of token blocks whose edges follow the 16-px grid. Both guided methods produce markedly narrower and stronger gradients at image structures. These include tooth outlines, the gingival margin and the lip as well as vessel borders and specular highlights of the placenta. Qualitatively, AnyUp yields the sharpest contours. Its edges are narrower than those of JBU and follow vessel borders and tooth outlines most closely. JBU depends more strongly on the bandwidth. With $\sigma_r = s_g/16$, selected on the ColorChecker, it also takes over texture, specular reflections and noise of the guide. With $\sigma_r = s_g/4$, the strong boundaries remain, but wider than with AnyUp. In low-contrast regions, JBU also traces the token grid as stripes. RGB and hyperspectral guides mostly yield similar edges. On textured skin, AnyUp takes over the skin texture with the RGB guide but hardly with the hyperspectral guide. The edge maps show where the CARL features change, not whether a true tissue boundary lies there. Appendix~\ref{app:medical} shows all images considered.

\subsection{CARL-Specific Analysis: Wavelengths and Band Sets}
CARL is camera-agnostic. It encodes each band together with its wavelength, allowing it to handle different band configurations. For users of the frozen encoder, it is therefore important to assess the required wavelength accuracy and whether guided upsampling remains effective across different band sets. We evaluate 78 configurations, including the reference setting of 32 bands spanning 400–820~nm, on both recordings. First, we examine how the features respond to wavelength shifts. We shift either the wavelength labels passed to CARL, the spectral bands sampled from the data or both together. Shifts range from $-50$ to $+50$~nm, depending on the available spectral range. Second, we investigate alternative band configurations, including 8 to 128 bands, random subsets, a restricted range of 400–700~nm and band positions corresponding to EnMAP and Sentinel-2. EnMAP~\cite{enmap} is a German hyperspectral satellite. We use the 69 centers of the EnMAP band grid from CARL's pretraining that lie in 400–820~nm (spacing 4.7–7.7~nm). Sentinel-2~\cite{sentinel2} is a multispectral satellite of ESA. Its seven bands B1–B7 lie in this range (Sentinel-2A, 443–783~nm). The hyperspectral guide of JBU always comes from the full cube and stays unchanged. The analysis thus tests robustness at a constant guide, not a real sensor that provides only these bands.

\begin{figure*}[!t]
\centering
\includegraphics{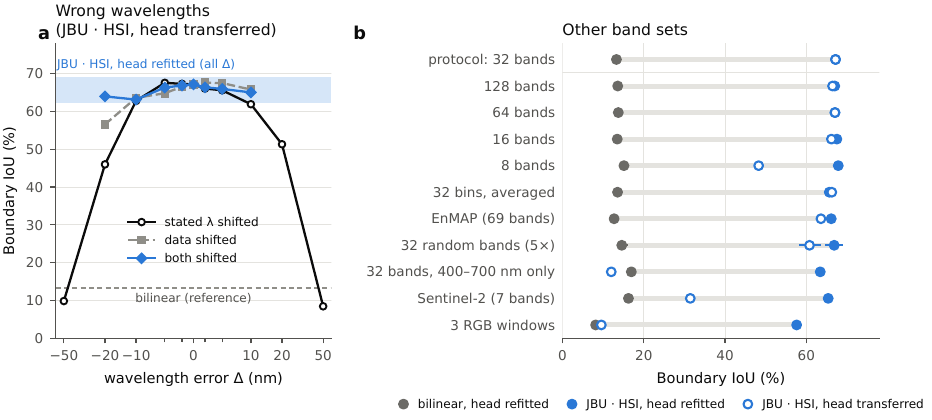}
\caption{Robustness to wavelengths and band sets (spatial CARL features, prototype head, mean of both recordings). (a) Boundary IoU after JBU with the transferred head when only the wavelengths passed to CARL (stated $\lambda$), only the data or both are shifted by $\Delta$. The data were shifted only up to +10~nm, with the top band limited by the cube end at 820.6~nm. Blue band: range with the refitted head. Dashed: bilinear. (b) Other band sets: bilinear (gray) and JBU (blue) with the refitted head. Open: JBU with the transferred head. Random bands: mean and range over five draws.}
\label{fig:wavelengths}
\end{figure*}

For each variation, we read out the features with the prototype head in two ways. The \emph{transferred} head was fitted once on the protocol setting and is reused unchanged. A user with a ready-made head would do the same. The \emph{refitted} head is determined on the features of the variation itself and shows whether the upsampler still helps independently of the head (Fig.~\ref{fig:wavelengths}).

With the transferred head, wavelength errors up to about $\pm 10$~nm are uncritical (at most 5~pp loss in Boundary IoU), larger errors are not. Wrongly stated wavelengths are the most harmful. If data and stated wavelengths are shifted together, the head remains largely usable. For other band sets, the spectral coverage matters more than the number of bands. If the range above 700~nm is missing, as for visible-only, Sentinel-2-like or RGB-like bands, the transferred head collapses. With the refitted head, in contrast, JBU outperforms bilinear interpolation by a median of 46 to 57~pp in all 78 variations and all 24 patches (Holm $p < 10^{-5}$). With changed bands, it is thus the transferred head that fails, not the upsampler.

\begin{figure*}[!t]
\centering
\includegraphics{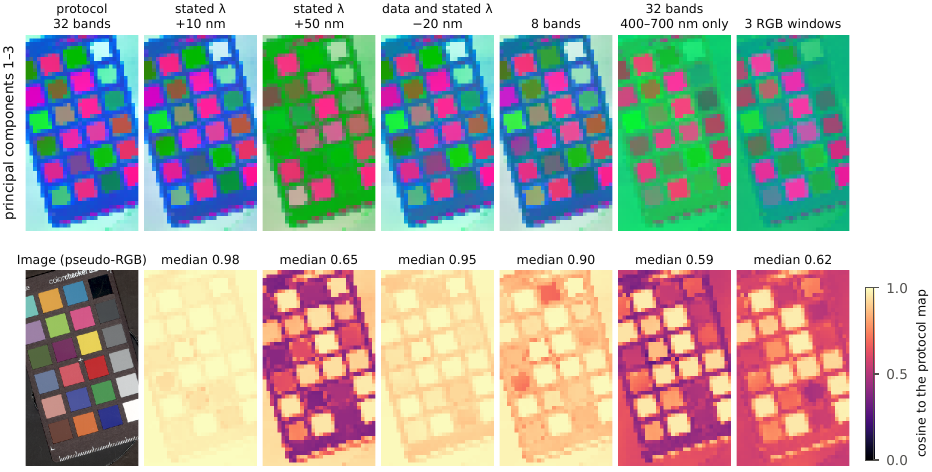}
\caption{Spatial CARL token maps of the rotated recording under selected band changes. Top: principal components 1–3, all in the color basis of the protocol map. Bottom: cosine of each token to the protocol map, with the median above. Stated $\lambda$: only the wavelengths passed to CARL shifted. Data and stated $\lambda$: both shifted. The maps of all 78 variations are published with the code.}
\label{fig:tokenmaps}
\end{figure*}

The token maps show where the features change (Fig.~\ref{fig:tokenmaps}). They remain nearly unchanged with a 10-nm error, with only eight bands or when data and stated wavelengths are shifted together (median cosine to the protocol map 0.90–0.98). With a 50-nm error or without the range above 700~nm, mainly the frame changes, whose spectrum is dark in the visible range and rises steeply above 700~nm. The patches remain internally uniform and distinguishable from each other. Because every patch edge borders the frame, the transferred head fails exactly at these edges, whereas a refitted head separates the classes again.

\section{Discussion}
Anyone who uses CARL without fine-tuning and reads out the features directly with a linear head or a prototype classifier loses a large part of the edge fidelity through bilinear interpolation. Image-guided upsampling recovers it without additional labels or further training. The two scalar parameters can be selected on a different recording. On the ColorChecker, JBU was the best or joint-best method in all endpoints, robust to the parameter choice and at the same time the most economical. AnyUp requires about twice the time and memory as well as a pretrained network. It also left frame contributions inside low-contrast patches. On CARL's token scale, the guided filter is a compromise between a sharp decision boundary and a clean patch interior.

Which method is preferable thus depends on the edges. Where boundaries are sharp and high-contrast, so that the bandwidth of JBU can be determined reliably on another recording, as on the ColorChecker, JBU is the most accurate method. Tissue has softer, lower-contrast and textured transitions. There, AnyUp qualitatively yields the sharper contours, presumably because its learned attention handles such transitions better than the fixed Gaussian kernel of JBU, which, depending on the bandwidth, takes over texture or widens edges (Fig.~\ref{fig:medical}). That AnyUp already shows the narrowest edge maps on the ColorChecker (Fig.~\ref{fig:teaser}b) but trails JBU quantitatively is consistent with its residual mixing inside dark patches, which an edge map does not show. For tissue, the choice therefore remains open until a quantitative evaluation is available. Whether sharper feature edges lead to more accurate segmentations there also depends on the head and on the annotation. Since the guide comes from the same recording, the camera independence of CARL is preserved in any case.

A hyperspectral guide helps where the visible contrast is small and the spectral difference is large. On the ColorChecker, this is rare. In tissue, we expect it more often, for example with differences in oxygenation or water content. The UV recording could not test this specifically, because the lamp does not change the reflectance of the chart. A real test requires materials that differ or fluoresce in the UV. It also requires a recording that preserves this and in which the UV bands are as sharp as the visible ones. The CARL-specific analysis further shows that band centers should be known to within about $\pm 10$~nm and that a head belongs to the band configuration on which it was fitted. The upsampler itself can be used unchanged with any configuration.

The study has limitations. The statistics refer to the 24 patches of one chart in two recordings. The third recording was evaluated only afterwards. With its sharp, high-contrast edges, the ColorChecker is a favorable case for any guide. The selected bandwidth lay at the edge of the pre-specified grid. The heads were fitted only on pure interior tokens, which isolates the effect of the upsampler but does not correspond to the usual training on all pixels. The band configurations are simulated from the same cubes, so the noise, spectral response and imaging properties of a real second camera are missing. The medical results are qualitative and based on a few images.

\emph{Next steps.} A natural next step is to test whether the gain holds once heads are trained on all pixels. Bilinear and guided upsampling could be compared under otherwise identical training, with a linear or a nonlinear head on frozen features and with fine-tuning of encoder and head. The upsampler could be applied either at inference only or already during training. A quantitative evaluation on medical datasets with dense annotations, for example in Boundary IoU and Boundary F1, would show whether the advantage carries over to segmentation.

\section{Conclusion}
Image-guided, training-free upsampling makes frozen CARL features markedly more edge-accurate. With JBU and the hyperspectral image as the guide, the Boundary IoU increases by 55~pp and the effective resolution of the features improves from about one token to a few pixels, without additional labels or further training. The gain replicates on a third recording, persists under wavelength errors and other band sets as long as the head matches the configuration and also appears qualitatively on tissue. For applications that use CARL with linear heads without fine-tuning, we recommend guided upsampling instead of bilinear interpolation, with JBU for sharp, high-contrast edges. On tissue with softer transitions, the qualitative results point to an advantage for AnyUp. Whether it holds must be shown by future quantitative studies. A natural next step is to examine whether the gain holds under training and fine-tuning for segmentation tasks or becomes marginal with nonlinear heads. The hyperspectral guide should also be tested on materials that differ only outside the visible range.

\section*{Data and Code Availability}
Code, result tables and notebooks are available at \url{https://gitlab.fb10.fh-dortmund.de/fb10-opensource/CARL-tokens-to-pixels}. The repository README provides access to the ColorChecker dataset with all acquisition metadata. A permanent DOI-based deposit of the dataset is in preparation. The placenta and ODSI-DB images are not redistributed. They are available from the original sources~\cite{placenta,odsi} under CC BY-NC-SA 4.0. The repository lists all images used. All experiments are available as executable scripts and notebooks. Apart from the runtime measurements, both execution paths yield byte-identical result tables.

\section*{Acknowledgment}
This research was funded by the German Federal Ministry of Research, Technology and Space (BMFTR) under the \emph{Forschung an HAW} program (grant number 13HAW021FQ). Equipment used in this study was purchased as part of a previous research project funded by the BMFTR (grant number 13FH097KX0).

\section*{Declaration of Competing Interest}
The authors declare that they have no known competing financial interests or personal relationships that could have appeared to influence the work reported in this paper.

\section*{Declaration of Generative AI and AI-Assisted Technologies in the Manuscript Preparation Process}
The authors conceived the study, acquired and prepared the ColorChecker data and conducted initial investigations, including comparisons of feature edge maps, Boundary IoU, effective resolution and simple qualitative tests of shifted wavelengths. ChatGPT (OpenAI, GPT-5.1) served as a sounding board for discussing initial ideas, visualizing first concepts and refining final text. Claude (Anthropic, Claude Opus 5.5 via Claude Code) was subsequently used to develop these investigations into a systematic and reproducible experimental framework with more readable, structured and consistent code. This included substantial generation and refinement of analysis and figure code, evaluation methods and statistical procedures, literature search assistance as well as drafting, translation and editing of manuscript text across all sections based on the authors' material. The contamination width $w_{90}$ was proposed by Claude in response to the authors' aim of quantifying contamination near boundaries. The AI-assisted development involved extensive iterative exchanges in which the authors specified requirements, critically evaluated suggestions, revised or rejected proposed approaches and requested additional checks. Throughout this process, the authors maintained the study’s scientific focus and ensured that the work formed a coherent, self-contained contribution. The authors also devoted substantial time and effort to iteratively reviewing and refining the manuscript and figures, with particular attention to readability and visual clarity. The authors defined the scope of the study, made the final scientific decisions and take full responsibility for the results and the manuscript.

\clearpage
\appendices
\setcounter{figure}{0}\renewcommand{\thefigure}{A\arabic{figure}}
\setcounter{table}{0}\renewcommand{\thetable}{A\arabic{table}}

\section{Edge Maps of All Methods}
\label{app:edgemaps}
\begin{figure*}[!t]
\centering
\includegraphics{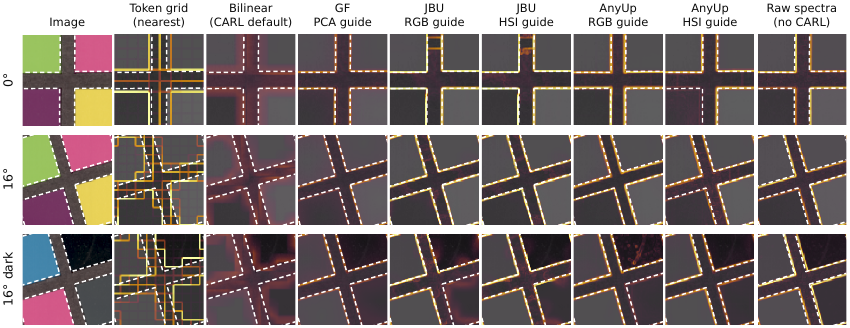}
\caption{Edge maps of all methods at three junctions (rows: upright, rotated, rotated with dark patches). Edge maps as in Fig.~\ref{fig:teaser}b, with one color scale for all CARL columns and a separate one for the raw spectra. White dashed: measured outline.}
\label{fig:A1}
\end{figure*}

Fig.~\ref{fig:A1} shows the edge maps of all upsamplers and of the raw spectra at three junctions. The pattern of Fig.~\ref{fig:teaser}b does not depend on the selection of methods or junctions. At the dark patches, the edge between the black patch and the frame is partly missing with the RGB guide but not with the hyperspectral guide.

\section{Where Does the Hyperspectral Guide Help?}
\label{app:h2}
\begin{figure*}[!t]
\centering
\includegraphics{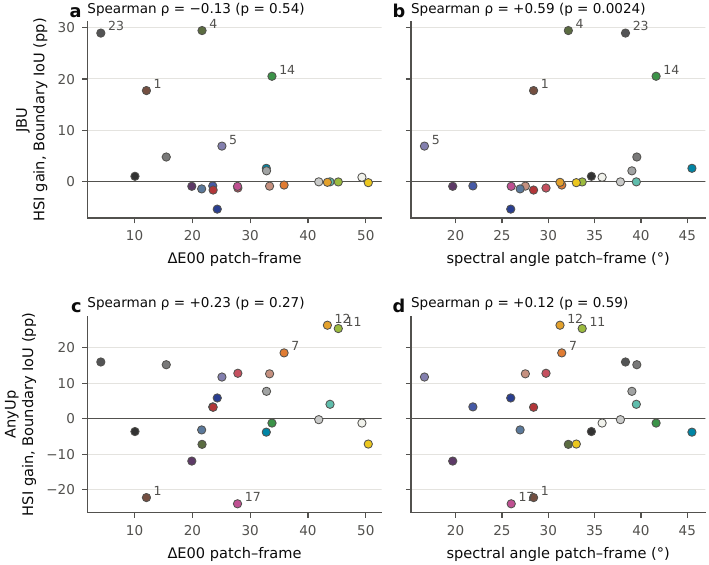}
\caption{Gain of the hyperspectral over the RGB guide in Boundary IoU per patch, top for JBU, bottom for AnyUp, versus the color difference $\Delta E_{00}$ (left) and the spectral angle between patch and frame (right). Each dot has the color of its patch. For each method, the five patches with the largest difference are numbered (1 Dark Skin, 4 Foliage, 5 Blue Flower, 7 Orange, 11 Yellow Green, 12 Orange Yellow, 14 Green, 17 Magenta, 23 Neutral 3.5).}
\label{fig:A2}
\end{figure*}

For JBU, the gain of the hyperspectral over the RGB guide is zero for most patches and large for a few. It follows the spectral angle between patch and frame (Spearman $\rho = 0.59$, $p = 0.002$), not the visible color difference $\Delta E_{00}$ ($\rho = -0.13$, Fig.~\ref{fig:A2}). For AnyUp, the gain in Boundary IoU scatters in both directions and depends on neither measure ($|\rho| \le 0.24$). In $w_{90}$, it follows the spectral angle ($\rho = 0.59$). The analysis was planned as descriptive. Table~\ref{tab:uv} summarizes the third recording with the UV lamp.

\begin{table}[!t]
\centering
\footnotesize
\caption{Third recording with the UV lamp switched on (spatial CARL features, prototype head, $\sigma_r = s_g/16$). IoU values in \%, mean over 24 patches. Median $w_{90}$ in native px. Bilinear uses no guide and serves as the reference for all rows. HSI: principal components of 400–820~nm. HSI+UV: of 325–820~nm. Added post hoc ($^\dagger$): UV only, with principal components of 325–400~nm. Half UV, with one half each from 400–820 and 325–400~nm.}
\label{tab:uv}
\begin{tabular}{lccc}
\toprule
Method & Boundary IoU & Corner IoU & $w_{90}$ (px) \\
\midrule
Bilinear & 15.6 & 10.7 & 34.5 \\
JBU, RGB & 65.0 & 76.5 & 3.5 \\
JBU, HSI & 69.3 & 81.2 & 3.5 \\
JBU, HSI+UV & 68.8 & 81.3 & 3.5 \\
JBU, UV only$^\dagger$ & 48.7 & 64.0 & 5.5 \\
JBU, half UV$^\dagger$ & 63.5 & 77.9 & 3.5 \\
AnyUp, RGB & 58.0 & 73.2 & 11.8 \\
AnyUp, HSI & 61.7 & 76.5 & 8.3 \\
AnyUp, HSI+UV & 61.5 & 76.4 & 8.3 \\
AnyUp, UV only$^\dagger$ & 33.0 & 44.0 & $\ge 100$ \\
AnyUp, half UV$^\dagger$ & 56.9 & 69.2 & 33.0 \\
\bottomrule
\end{tabular}
\end{table}

\section{Parameter Sensitivity}
\label{app:sens}
\begin{figure*}[!t]
\centering
\includegraphics{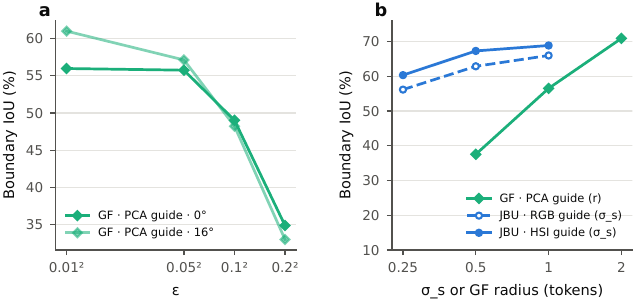}
\caption{Boundary IoU (prototype head) as a function of (a) $\varepsilon$ for the guided filter in both recordings and (b) the fixed parameters $\sigma_s$ (JBU) and filter radius (guided filter).}
\label{fig:A3}
\end{figure*}

For the guided filter, the Boundary IoU decreases with $\varepsilon$. The selection across recordings misses the optimum in the rotated recording by 3.9~pp. A larger filter radius sharpens the decision boundary but carries residual mixing deeper into the patch (Fig.~\ref{fig:A3}).

\section{Edge Measures beyond $w_{90}$}
\label{app:edge}
\begin{figure*}[!t]
\centering
\includegraphics{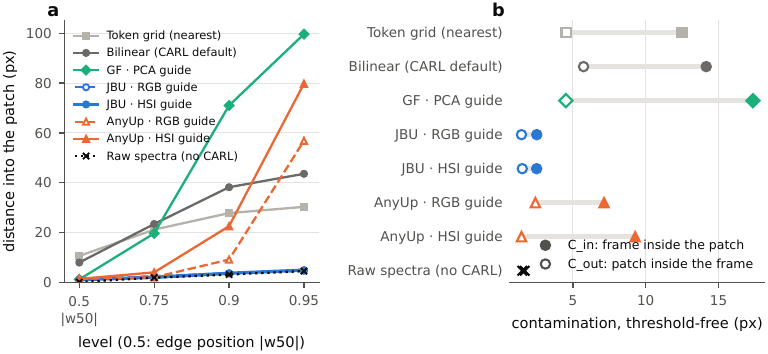}
\caption{(a) Distance into the patch beyond which the apparent patch fraction $q$ permanently reaches the respective threshold (median over patches, mean of both recordings). At 0.5, this is the absolute edge position. (b) Threshold-free contamination: frame remaining inside the patch ($C_{\mathrm{in}}$, filled) and patch leaking into the frame ($C_{\mathrm{out}}$, open).}
\label{fig:A4}
\end{figure*}

Because the threshold of 0.9 is arbitrary and $w_{90}$ conflates edge position and residual mixing, we computed after the main analysis, from the same profiles, the edge position $w_{50}$, the widths $w_{75}$ and $w_{95}$ as well as two threshold-free integrals, the frame remaining inside the patch and the patch leaking into the frame,
\begin{equation}
C_{\mathrm{in}} = \int_0^{L} \lvert 1 - q(s) \rvert \, ds, \qquad C_{\mathrm{out}} = \int_{-L'}^{0} \lvert q(s) \rvert \, ds.
\end{equation}
Compared with bilinear interpolation, JBU improves all these measures in 24 of 24 patches (Holm $p < 0.001$). The guided filter places the edge correctly ($|w_{50}| \approx 1$~px) but leaves more frame inside the patch than bilinear. AnyUp is on par with JBU in edge position and $w_{75}$. Its residual mixing appears only from the threshold of 0.9 onward (Fig.~\ref{fig:A4}). These measures were not pre-registered.

\section{Additional Tables}
\label{app:tables}

\begin{table*}[!t]
\centering
\footnotesize
\caption{As Table~\ref{tab:main}, for the spectral CARL feature map.}
\label{tab:spectral}
\begin{tabular}{lccccc}
\toprule
Method & Boundary IoU (prototype) & Boundary IoU (linear) & Corner IoU (prototype) & Patch IoU (prototype) & $w_{90}$ (px) \\
\midrule
Token grid (nearest) & 11.5 & 9.8 & 24.8 & 76.8 & 24.9 \\
Bilinear (CARL default) & 22.4 & 16.8 & 15.0 & 84.7 & 34.4 \\
Guided filter, PCA guide & 67.9 & 58.8 & 71.0 & 96.1 & 58.8 \\
JBU, RGB guide & 71.2 & 62.5 & 83.7 & 96.7 & 3.6 \\
JBU, HSI guide & 75.8 & 66.9 & 88.5 & 97.4 & 3.6 \\
AnyUp, RGB guide & 65.3 & 61.8 & 80.9 & 96.2 & 5.2 \\
AnyUp, HSI guide & 69.8 & 63.0 & 84.4 & 95.7 & 7.1 \\
\bottomrule
\end{tabular}
\end{table*}

\begin{table}[!t]
\centering
\footnotesize
\caption{Boundary IoU (\%) with heads from the same or the other recording (spatial features, pooled over both recordings).}
\label{tab:cross}
\begin{tabular}{lcccc}
\toprule
& \multicolumn{2}{c}{Prototype} & \multicolumn{2}{c}{Linear} \\
\cmidrule(lr){2-3} \cmidrule(lr){4-5}
Method & same & other & same & other \\
\midrule
Token grid & 8.9 & 8.8 & 9.9 & 9.6 \\
Bilinear & 13.3 & 12.8 & 18.5 & 17.2 \\
Guided filter & 56.5 & 53.2 & 66.4 & 62.1 \\
JBU, RGB & 62.9 & 61.7 & 67.1 & 64.3 \\
JBU, HSI & 67.3 & 65.3 & 70.3 & 67.8 \\
AnyUp, RGB & 60.5 & 57.6 & 64.3 & 63.1 \\
AnyUp, HSI & 63.7 & 61.2 & 69.2 & 65.3 \\
\bottomrule
\end{tabular}
\end{table}

\begin{table}[!t]
\centering
\footnotesize
\caption{Runtime and GPU memory per image for upsampling one feature map ($768 \times 52 \times 30$ to $520 \times 300$~px, NVIDIA H100, median of five runs), excluding the computation of the guide. PC: principal component of the hyperspectral image.}
\label{tab:runtime}
\begin{tabular}{lcc}
\toprule
Method (guide) & Time (ms) & Memory (MB) \\
\midrule
Token grid (nearest) & 0.4 & 462 \\
Bilinear (CARL default) & 0.5 & 457 \\
Guided filter (PC 1–3) & 115 & 7326 \\
JBU (RGB or 16 PCs) & 40 & 1408 \\
AnyUp (RGB or PC 1–3) & 90 & 2812 \\
\bottomrule
\end{tabular}
\end{table}

\section{Synthetic Edge Test}
\label{app:synthetic}
As a plausibility check of the implementation and of $w_{90}$, we use an ideal step between two materials that does not lie on a token boundary. The tokens are exact footprint means and the guide is the ideal step. This yields $w_{90}$ values of 18~px (nearest), 34~px (bilinear), 46~px (guided filter), 3~px (JBU) and 7~px (AnyUp). With a perfect guide, JBU and AnyUp nearly reach the resolution of the evaluation grid, whereas the guided filter does not, which explains its tail on the real data.

\section{All Medical Images Considered}
\label{app:medical}
\begin{figure*}[!t]
\centering
\includegraphics{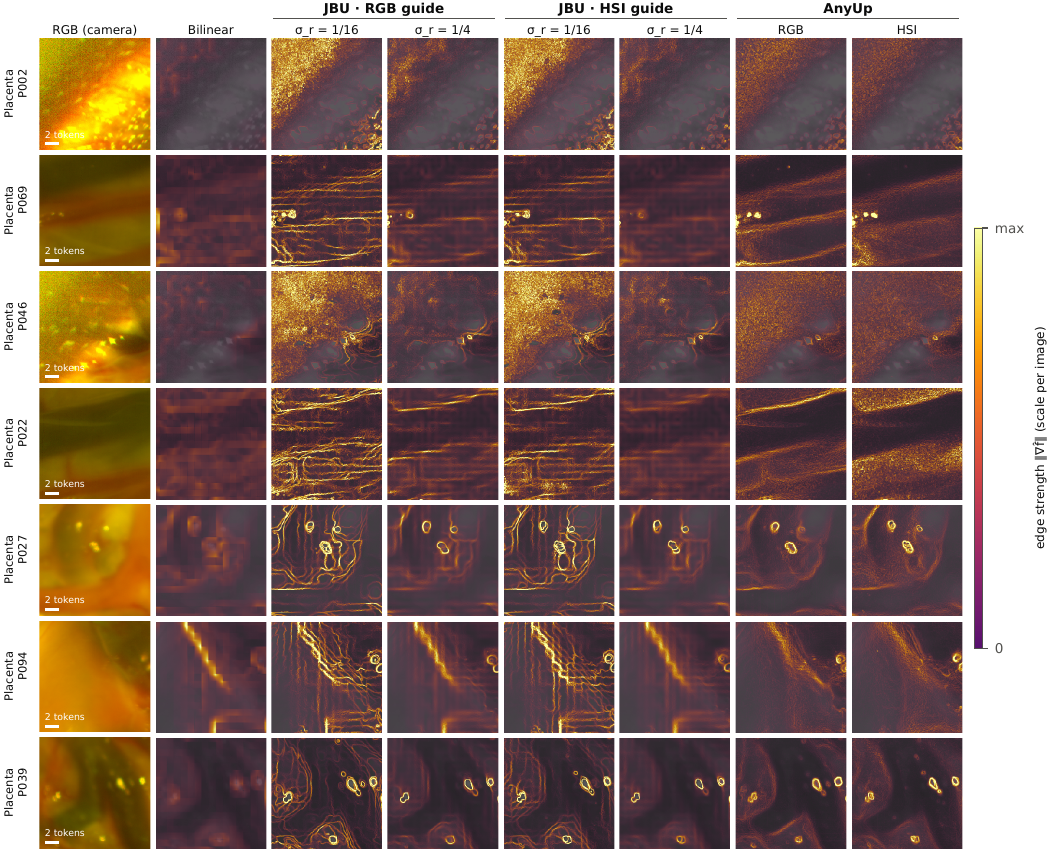}
\caption{All placenta candidates, displayed as in Fig.~\ref{fig:medical}. For P002 and P046, the crop lies in the image corner because all other windows contain saturated specular highlights. Images adapted from~\protect\cite{placenta}. Licensed under CC BY-NC-SA 4.0 and not covered by the license of this article.}
\label{fig:A5}
\end{figure*}

\begin{figure*}[!t]
\centering
\includegraphics{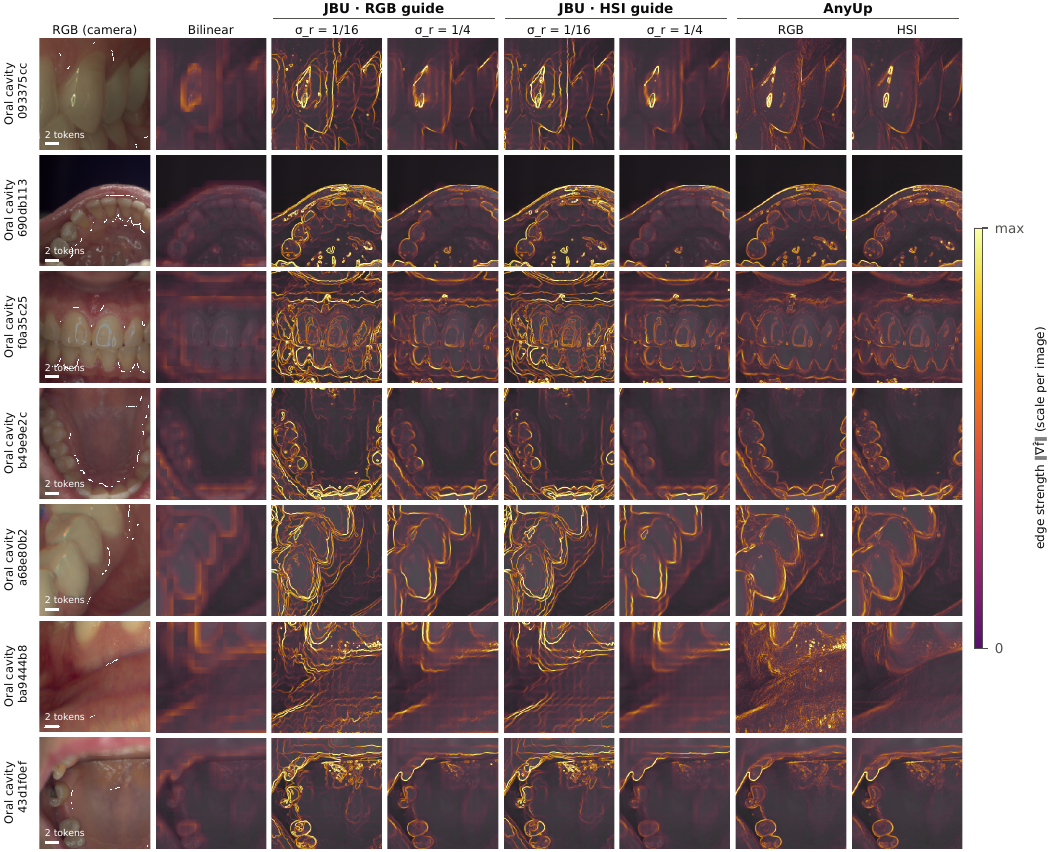}
\caption{All oral cavity candidates from ODSI-DB~\protect\cite{odsi}, displayed as in Fig.~\ref{fig:medical}. Images adapted. Licensed under CC BY-NC-SA 4.0 and not covered by the license of this article.}
\label{fig:A6}
\end{figure*}

Figs.~\ref{fig:A5} and~\ref{fig:A6} show all candidates from which the images in Fig.~\ref{fig:medical} were chosen. For each acquisition session, this is the image with the longest annotated class boundaries. For the placenta, all six sessions and one additional image are shown. For the oral cavity, the six sessions with the longest class boundaries and one image of the posterior teeth are shown. The crop is in each case the window with the densest annotated boundaries, on the placenta without saturated specular highlights. The pattern of Fig.~\ref{fig:medical} appears in all images. In strongly textured or noisy regions, all guided methods also follow the noise of the guide.

\end{document}